\documentclass{aa}  
\usepackage[table, rgb, dvipsnames]{xcolor} 
\usepackage{graphicx}
\usepackage{txfonts}
\usepackage{float}
\usepackage{tikz}
\usepackage{upgreek}

\begin{document}

   \title{To collapse or not to collapse: Halo evolution with self-interacting dark matter mass segregation}
   \titlerunning{SIDM mass segregation}
   \author{Yashraj Patil
           \inst{\ref{inst:usm}, \ref{inst:bits}, \ref{inst:umn}} and
           Moritz S.\ Fischer\inst{\ref{inst:usm},\ref{inst:origins}}
          }
    \authorrunning{Y.\ Patil \& M.\ S.\ Fischer}

    \institute{
        Universitäts-Sternwarte, Fakultät für Physik, Ludwig-Maximilians-Universität München, Scheinerstr.\ 1, D-81679 München, Germany\label{inst:usm}\\
        \email{ypatil@usm.lmu.de}\\
        \email{mfischer@usm.lmu.de}
        \and
        Birla Institute of Technology and Science-Pilani, K. K. Birla Goa campus, NH-17B, Zuarinagar, Goa 403726, India\label{inst:bits}
        \and
        College of Science and Engineering, University of Minnesota, 117 Pleasant St, Minneapolis, MN 55455, USA\label{inst:umn}
        \and
        Excellence Cluster ORIGINS, Boltzmannstrasse 2, D-85748 Garching, Germany\label{inst:origins}
    }

   \date{Received 30 June, 2025 / Accepted 3 September, 2026}

  \abstract
   {Surprisingly compact substructures in galaxies and galaxy clusters, as well as field halos, observed by gravitational lensing could be challenging to explain solely on the basis of collisionless dark matter (DM). To investigate such objects, recent studies have focussed on the gravothermal collapse that occurs in halos consisting of self-interacting dark matter (SIDM). However, simple models of elastic scattering could face issues when attempting to explain the nature of these compact objects during very late stages of the collapse and the post-collapse phase, where a black hole may have formed from DM.}
   {We aim to explain compact halos, while avoiding the gravothermal catastrophe that conventional SIDM models are subject to. We investigate the evolution of a DM halo for a model consisting of two species with unequal masses, which features only non-gravitational interactions among the different species, but not within their own structures.}
   {By employing $N$-body simulations, we studied
   the effect of unequal-mass scattering on the evolution of an isolated DM halo. In particular, we simulated the late stage of its evolution with a high central density.}
   {We find that our two-species models can produce density cores with sizes that are dependent on the mass ratio between the two species. Moreover, mass segregation caused by the unequal particle masses leads to an enhanced finite final central density, or at least a slowly growing state, which depends on the mass ratio and the mass fraction of the two DM species.}
   {We find that SIDM models consisting of two DM species have the capacity to simultaneously explain DM halos with density cores and systems that are denser in their centre than would otherwise be expected from collisionless DM, while avoiding the gravothermal catastrophe scenario. This approach offers a compelling alternative to single-species models, along with a rich phenomenology.}


   \maketitle

\section{Introduction}
    Lambda cold dark matter ($\Lambda$CDM) has long remained the standard framework for investigating dark matter (DM), with results that offer a good agreement with observations on cosmological scales, based on the large-scale structure. However, little is known about the nature of DM and a viable particle candidate has yet to be detected in the laboratory \citep{Cirelli_2024}. Although the CDM model works quite well to make successful predictions on large scales, it faces challenges on smaller scales; specifically, when attempting to explain the diversity of observed galactic rotation curves \citep[e.g.][]{Oman_2015, Sales_2022}.

    Recent observations of surprisingly dense substructures are challenging to explain with collisionless cold DM \citep{Vegetti_2010, Meneghetti_2020, Minor_2021, Granata_2022, Despali_2025a, Minor_2025, Cao_2026, Vegetti_2026}. Overall, CDM does predict fairly compact substructure, but with the caveat that it is very rare \citep{Tajalli_2025}.\footnote{The lensing subhalo in SDSSJ0946+1006 may also be explained by a less dense substructure when allowing for a subhalo that is not completely dark, which makes it more common within $\Lambda$CDM \citep{He_2025}.} Given a high abundance of observed dense substructures compared to CDM simulations, this might call for an alternative mechanism behind the formation of these objects. In recent studies, alternative DM models have been investigated to study structure formation on galactic scales and so, self-interacting dark matter (SIDM) stands out as a promising class of models. It has the potential to address discrepancies between $\Lambda$CDM predictions and observations. The rich phenomenology of SIDM has been intensively studied in recent years \citep[e.g.][]{Rose_2022, Correa_2024, Moreno_2025, Ragagnin_2024, Despali_2025b, Straight_2025}. A comprehensive overview of SIDM is provided in the review articles by \cite{Tulin_2018, Adhikari_2025}.

   Overall, SIDM can involve elastic scattering between DM particles along with their gravitational interactions. The energy transfer resulting from these interactions shapes the DM halos during their gravothermal evolution. Such a model proves to be a potential candidate to replace CDM as it is promising in replicating the observed range of central densities in DM halos \citep{Nadler_2023}. Recent studies have proposed the gravothermal collapse of SIDM halos as a possible mechanism to form highly concentrated halos to explain the high central densities \citep[e.g.][and the references therein]{Turner_2021, Yang_2023, Li_2025, Zeng_2025, Fischer_2026a, Yu_2026}. This could give rise to an observational signature that allows it to be discerned from a collisionless model.

    In this paper, we employ a DM model that consists of two particle species with different masses. Importantly, non-gravitational DM interactions occur only between particles of different species, but not within any of the two species. This means that a particle belonging to the lighter species scatters with particles of the heavier species, but not with particles of the lighter species. At the same time, there are no interactions allowed between particles of the heavier species. A similar model was recently studied with cosmological $N$-body simulations by \cite{Yang_2025, Yang_2026}. Our asymmetric DM scenario is motivated by a DM number-violating Majorana-type mass that can give rise to particle-antiparticle oscillations, as studied by \cite{Tulin_2012}. We note that for the purposes of this study, we only consider elastic scattering.

    Our focus lies in exploring the possibility of simulating such a model and understanding its impact on the evolution of an isolated halo. The unequal particle masses can give rise to mass segregation, which is a mechanism that is known to possibly change the halo structure. Previously, this has been investigated for DM being partially made of primordial black holes \citep[e.g.][]{Boldrini_2020} and was considered in the context of SIDM by \cite{Patil_2024, Yang_2025}. Such a two-species model is particularly interesting as it allows for more compact halos with an increased central density. The typical formation scenario for such a surprisingly compact object in a single-species model is the gravothermal collapse of an SIDM halo. However, with our model, we present an alternative formation mechanism based on mass segregation with interesting consequences in terms of the late-time evolution of the DM halo.
    
    We find that our two-species model can allow for core expansion similar to the commonly investigated single-species models. Interestingly, in the late phases of the evolution, it can drastically reduce the growth of the central density, effectively avoiding a gravothermal catastrophe as well as the potentially related formation of a black hole (BH) from DM at the centre of the halo \citep[e.g.][]{Feng_2022, Feng_2025}.

    The remainder of the paper is structured as follows. In Sect.~\ref{sec:model}, we describe the DM model of this study in detail and explain its expected impact on the evolution of DM halos. Subsequently, we describe in Sect.~\ref{sec:method} how we can simulate this model with the help of our $N$-body code \textsc{OpenGadget3}, including the details of the implementation. A test problem to verify the numerical scheme and its implementation is presented in Sect.~\ref{sec:tests}. In Sect.~\ref{sec:isolated_halo}, we study the effect of our DM model on the evolution of an isolated halo. Limitations of our work and physical implications for DM, including the advantages of a two-species model compared to the limitations of a single-species model, as well as the possibility of constraining such a two-species DM model, are discussed in Sect.~\ref{sec:discussion}. Finally, in Sect.~\ref{sec:conclusion}, we summarise and present our conclusions. Additional information is provided in the appendices.

\section{SIDM with two species} \label{sec:model}
\begin{figure}
    \centering
    \begin{tikzpicture}
        \node[anchor=north west] (A) at (-1,0) {
            \includegraphics[width=0.41\columnwidth]{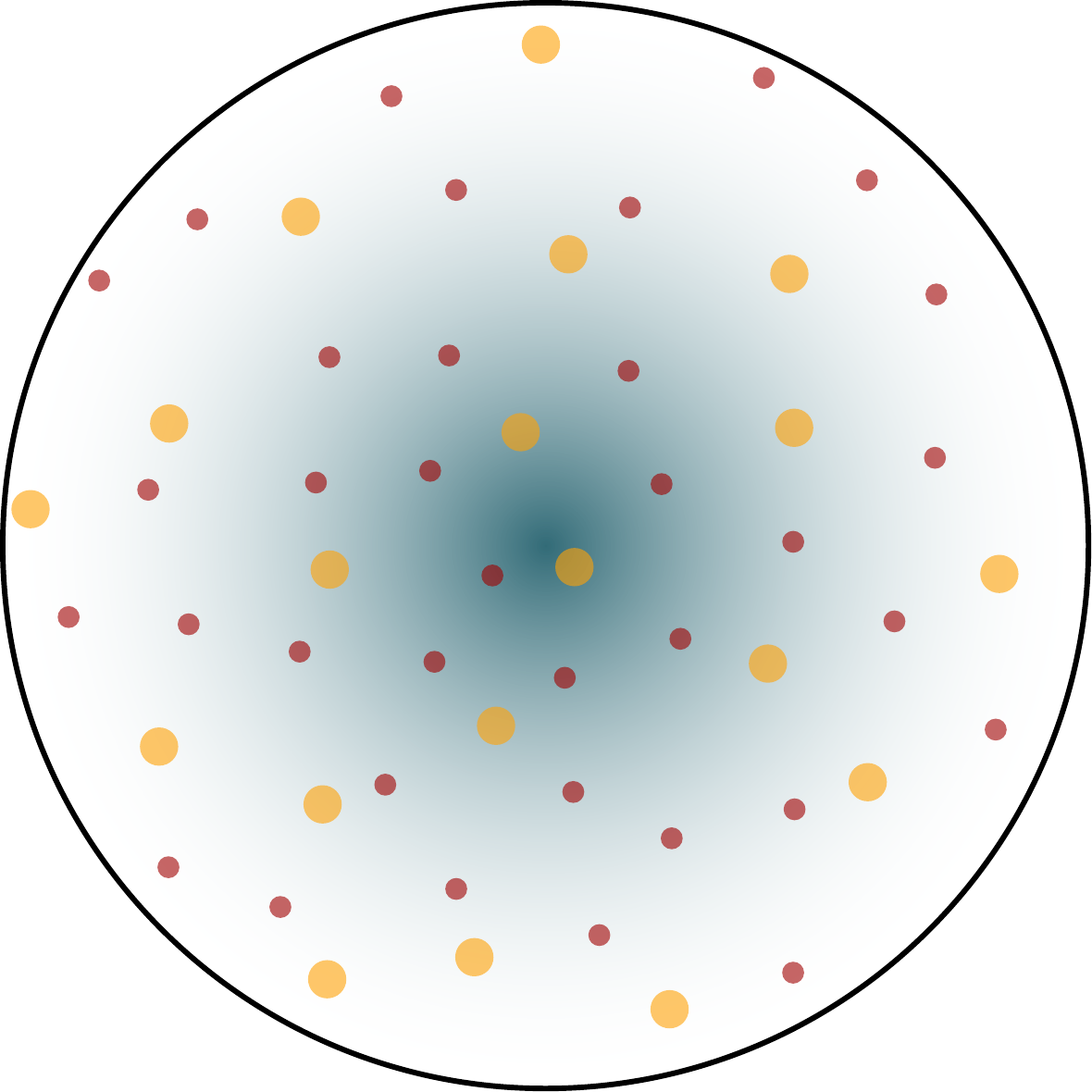}
        };

        \node[anchor=north west] (B) at (4.2,0) {
            \includegraphics[width=0.41\columnwidth]{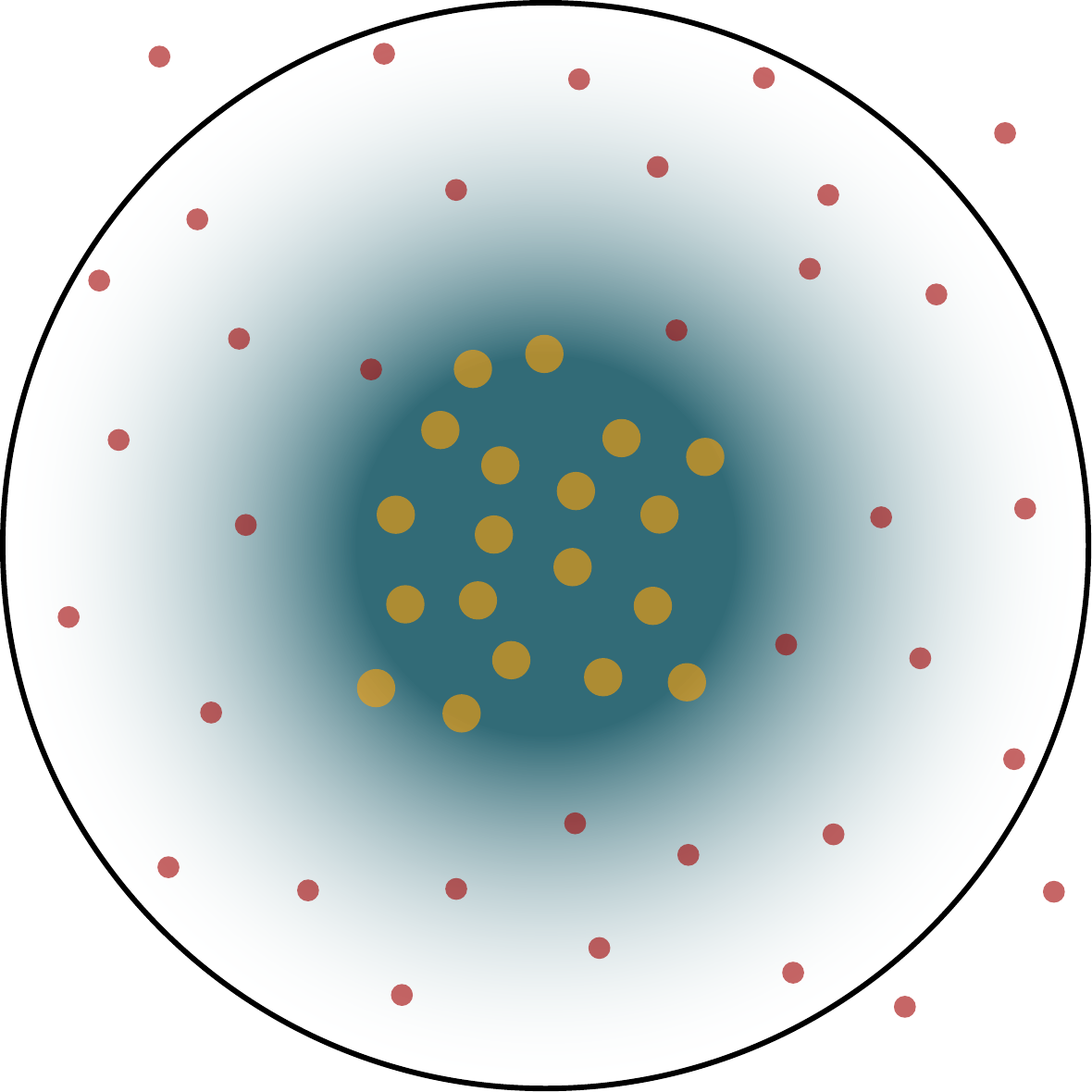}
        };

        \draw[->, thick] (A.east) -- (B.west)
            node[midway, above, font=\small]
            {\shortstack{Time \\ Evolution}};
    \end{tikzpicture}

    \vspace{-2mm}

    \begin{tikzpicture}
        \node[anchor=north west] (C) at (1,0) {
            \includegraphics[width=0.50\columnwidth]{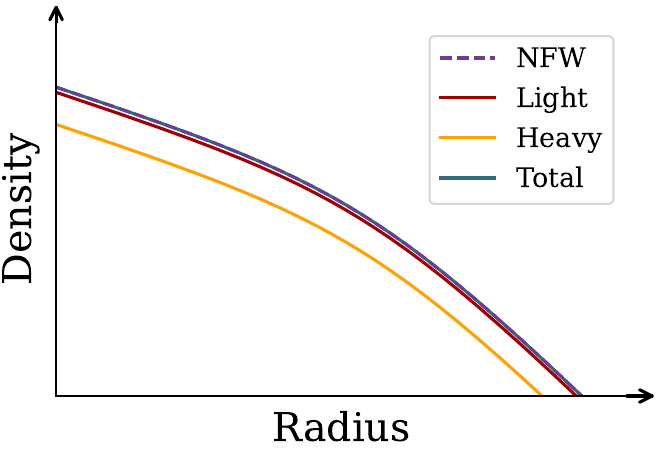}
        };

        \node[anchor=north west] (D) at (5.3,0) {
            \includegraphics[width=0.50\columnwidth]{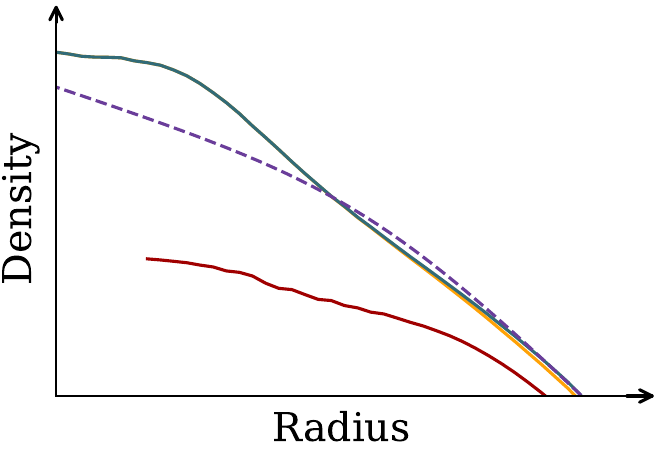}
        };
    \end{tikzpicture}
    
    \caption{Illustration of the effect of dynamical mass segregation in an isolated SIDM halo with two DM species. \textit{Left column}: Initial state. \textit{Right column}: Final evolved state. In the upper panels, the larger golden dots represent the heavier $\mathrm{{DM}_{h}}$ particles and the smaller red dots represent the lighter $\mathrm{{DM}_{l}}$ particles. Initially, the distribution of both species is assumed to be uniform throughout the halo (top-left), as seen from the density profile (bottom-left). After a certain time has passed, dynamical mass segregation causes the heavier particles to sink into the centre, while the lighter particles move to outer orbits (top-right). During this process, some particles escape the halo's gravity due to high velocities, represented by dots outside the halo. The density profile for $\mathrm{{DM}_{h}}$ shows an increased density in the central region, contrary to the $\mathrm{{DM}_{l}}$ profile (lower-right panel). Moreover, with the spatial separation of the two species, the halo evolves towards a state with a stable and higher central density in the halo, avoiding gravothermal collapse.}
    \label{fig:time_evolution}
\end{figure}

Our SIDM model of interest consists of two DM species differentiated by their mass, the heavier DM species ($\mathrm{DM_{h}}$) and the lighter DM species ($\mathrm{DM_{l}}$). The non-gravitational interactions only take place between particles of different species and not between particles of the same species.\footnote{Although there are no self-interactions within one of the two species, but only inter-species interactions, we consider this model as belonging to the class of SIDM models. In this sense, `self' does not refer to a specific species but only means that particles that make up the DM interact with each other, in contrast to interactions with baryons.} This is motivated by a DM number-violating Majorana-type mass \citep{Tulin_2012}.

To specify the strength of the interactions, we give the cross-section divided by the physical particle mass of the lighter species, ${\sigma}/{m_\mathrm{DM_{l}}}$. The total cross-section, $\sigma$, is proportional to the number of scattering events,
\begin{equation} \label{eq:total_cross_section}
    \sigma = \int^1_{-1} \frac{\mathrm{d} \sigma}{\mathrm{d} \cos \theta_\mathrm{cms}} \, \mathrm{d} \cos \theta_\mathrm{cms} \,.
\end{equation}
Here, `cms' refers to the centre-of-mass frame of the two scattering particles.

However, the energy and momentum transfer between the two species depend on the momentum transfer cross-section \citep[e.g.][]{Dvorkin_2014}, expressed as
\begin{equation} \label{eq:momentum_transfer_cross_section}
    \sigma_\mathrm{T} = \int_{-1}^{1} \frac{\mathrm{d} \sigma}{\mathrm{d} \cos \theta_{\mathrm{cms}}}\left(1-\cos \theta_{\mathrm{cms}} \right) \mathrm{d} \cos \theta_{\mathrm{cms}}\, .
\end{equation}
We commonly use this equation to specify the strength of the interactions. This is in contrast to a single-species model, where the angle-averaged cross-section that is typically better suited to specify the strength of the interaction is the viscosity cross-section \citep[e.g.][]{Yang_2022D, Sabarish_2024}. We used it in this study as well and defined it as
\begin{equation} \label{eq:viscosity_cross_section}
\sigma_\mathrm{V} = \int_{-1}^{1} \frac{\mathrm{d} \sigma}{\mathrm{d} \cos \theta_{\mathrm{cms}}}\sin^2\theta_{\mathrm{cms}} \, \mathrm{d} \cos \theta_{\mathrm{cms}}
\, .
\end{equation}

In addition, we introduced two new parameters: a mass fraction, $f$, and mass ratio, $\mu$, to specify the properties of the DM models we simulated. The mass fraction is defined as: $f = \mathrm{M_{DM_l} / M_{total}}$ where $\mathrm{M_{DM_l}}$ is the total mass of the lighter species and $\mu$ is defined as: $\mu = {m_\mathrm{{DM}_{h}}}/{m_\mathrm{{DM}_{l}}}$, where ${m_{\mathrm{DM}_{x}}}$ is the mass of the physical particle of species $x$. 

In this study, we focus on the dynamical mass segregation that occurs in gravitationally bound systems \citep{Spitzer_1969, Allison_2009, Olczak_2011}, originating from the energy exchange resulting from self-interactions between the two species. During these encounters, the redistribution of energy tends towards the velocities of each species following a Maxwell-Boltzmann distribution, resulting in the average energy per particle approaching the same value for all species, establishing energy equipartition.

In the case of a DM halo, when assuming that both species initially follow the same velocity distribution, the interactions have a distinct impact. They make the heavier DM particles lose energy through scattering to the lighter DM particles, causing them to move more slowly than the lighter ones. Consequently, the $\mathrm{DM_{h}}$ particles move to the lower orbits and effectively sink towards the centre of the DM halo, while the $\mathrm{DM_{l}}$ particles travel outward to higher orbits. As the system evolves, the late stages of the halo evolution show a more concentrated distribution of $\mathrm{DM_{h}}$ particles near the centre and a broader distribution of $\mathrm{DM_{l}}$ towards the outer regions, as shown in Fig.~\ref{fig:time_evolution}. This process of mass segregation leads to an energy outflow and mass inflow, increasing the central density. The final central density observed is significantly higher compared to the initial and since very few lighter particles that can carry energy to the outer orbits remain, the rise slows down and becomes steady after a certain time. In this way, we were able to avoid the gravothermal collapse phase, as observed in single-species SIDM scenarios.

With our simulations, we studied models with mass ratios in the range of $\mu \in [1,4]$ and a mass fraction of the lighter species of $f \in [0.05, 0.5]$. In principle, models lying outside of this range could be of interest too, but they are beyond the focus of this paper. For the sake of simplicity, we only considered a velocity-independent cross-section for the interactions. Moreover, we investigated two different angular dependencies. In Sect.~\ref{sec:tests}, we drew a comparison between the evolution time scale for isotropic and forward-dominated scenarios using a simple test set-up. Subsequently, in Sect.~\ref{sec:isolated_halo}, we only considered forward-dominated scenarios for the evolution of an isolated DM halo. 

\section{Numerical method} \label{sec:method}
In this section, we describe the numerical setup used for our work. We detail our modifications to an existing SIDM implementation with an aim to model self-interactions between two different species of DM particles and our choice of the cross-sections used in the simulations.

We used the cosmological $N$-body simulation code \textsc{OpenGadget3} \citep[e.g.][]{Groth_2023, Dolag_2026}, which is a successor of \textsc{Gadget-2} \citep{Springel_2005}. The domain decomposition and the neighbour search we employed were described by \cite{Ragagnin_2016}. The code contains a module for SIDM developed by \cite{Fischer_2021a, Fischer_2021b, Fischer_2022, Fischer_2024a, Fischer_2026b}. This SIDM module is capable of simulating two regimes for self-interactions based on their scattering angle: frequent self-interactions (fSIDM) and rare self-interactions (rSIDM). In particular, fSIDM corresponds to the limit of a very anisotropic cross-section where the size of scattering angles approaches zero, while the number of interactions increases as the momentum transfer cross-section is kept constant. The differential cross-section in this case can be given by 
\begin{equation}    \label{eq:diff_eq}
    \left.\frac{\mathrm{d}\sigma}{\mathrm{d}\cos\theta_\mathrm{cms}}\right|_\mathrm{fwd}=\lim_{\epsilon\to 0} \frac{\sigma_\mathrm{T}}{4 \, \ln(\epsilon^{-2})} \frac{1}{\left(\epsilon^2 + \sin^2{\theta_\mathrm{cms}/2}\right)^2}\,,
\end{equation}
where the subscript `cms' stands for quantities measured in the centre-of-mass system of the two scattering particles. Here, $\sigma_\mathrm{T}$ is the transfer cross-section given in Eq.~\eqref{eq:momentum_transfer_cross_section}, $\theta_\mathrm{cms}$ is the angle of scattering. Such a differential cross-section can be effectively described by a drag force as in \cite{Kahlhoefer_2014} and perpendicular momentum diffusion \citep{Fischer_2021a}. A detailed discussion of the validity of the fSIDM description for various differential cross-sections is given by \cite{Arido_2025}. To take an unequal mass ratio into account, we modified the drag force equation for the pairwise interactions between two different numerical DM particles $i$ and $j$ as
\begin{align} \label{eq:f_drag_new}
    F_{\mathrm{drag}} =& \, |\Delta\mathbf{v}_{ij}|^2 \, \frac{\sigma_\mathrm{T}}{m_\mathrm{{DM}_{l}}} \,  \frac{1}{1+\mu}\,m_\mathrm{n,h}  \,m_\mathrm{n,l} \Lambda_{ij} \,.
\end{align}
Here, $\Delta\mathbf{v}_{ij}$ is the relative velocity of particles $i$ and $j$, $m_\mathrm{{DM}_{x}}$ denotes the physical DM particle mass, and $m_{\mathrm{n},x}$ denotes the numerical DM particle mass where `l' stands for the lighter DM particle and `h' stands for the heavier DM particle. Then, $\Lambda_{ij}$ is the kernel overlap integral where we employed the spline kernel introduced by \cite{Monaghan_1985}. We calculated the kernel overlap as detailed in \cite{Fischer_2021a}, given as
\begin{equation} \label{eq: Lambda_ij}
    \Lambda_{ij} = \int W(|\mathbf{x}-\mathbf{x}_i|,h_i) \,  W(|\mathbf{x}-\mathbf{x}_j |,h_j) \, \mathrm{d}^3\mathbf{x} \, .
\end{equation}
The kernel size, $h$, was chosen adaptively to enclose a certain number of nearest neighbour particles in our case 48. The momentum transfer cross-section, $\sigma_\mathrm{T}$, is similar to the one used for single-species models with indistinguishable particles; as, for example, done by \cite{Kahlhoefer_2014, Robertson_2017b}. However, for our case of distinguishable particles, we used Eq.~\eqref{eq:momentum_transfer_cross_section}, which is not symmetric about a scattering angle of $\uppi / 2$. Our implementation also requires the mass ratio of numerical particles to be the same as that of the physical particles to achieve the explicit conservation of linear momentum.

Rare self-interactions (i.e.\ large-angle scatterings) can be modelled by computing the scattering probability the same way as in \cite{Rocha_2013a}; namely, by employing the kernel overlap, $\Lambda_{ij}$ \citep[see eq.~13 by][]{Fischer_2021a}. We built upon the existing rare self-interaction implementation by \cite{Fischer_2021a}, making slight adjustments in the equation to accommodate for the unequal masses of the relevant species. Following the derivation in \cite{Fischer_2021a}, we obtained the scattering probability for particles $i$ and $j$, given as
\begin{align} \label{eq:interact_prob}
    P_{ij} = \frac{\sigma}{m_\mathrm{DM_{l}}} \, m_\mathrm{n,l} \, \Delta v_{ij} \, \Delta t \, \Lambda_{ij}.
\end{align}
Here, $\sigma$ denotes the total cross-section (Eq.~\ref{eq:total_cross_section}).
Moreover, $\Delta t$ refers to the time-step explained further. In our simulations, when the gravitational time-step becomes larger than what is needed for scattering, we employed a time-step criterion for velocity-independent self-interactions similar to the one introduced in \cite{Fischer_2022, Fischer_2024a}. We computed the fractional velocity change with respect to the numerical particle in the centre-of-mass frame. Using the maximum allowed fractional velocity change, $\tau$, we can express the time-step criterion as
\begin{align}
    \Delta t_i <& \, \tau \, \frac{1}{v_\mathrm{max}\,{{m_\mathrm{n,l}}}\Lambda_{ii}} \, \left(\frac{\sigma_\mathrm{T}}{m_\mathrm{{DM}_{l}}}\right)^{-1} \, ,
\end{align}
where $v_\mathrm{max}$ is the maximum relative velocity that a particle experienced in the previous time step and $\Lambda_{ii}$ is the kernel overlap of the particle with itself. In the case of rSIDM, we would need to replace $\sigma_\mathrm{T}$ with $\sigma$. We note that Eqs.~\eqref{eq:f_drag_new} and~\eqref{eq:interact_prob} are specific versions of the more general equations given by \cite{Fischer_2025a}. For a detailed derivation, we refer to that paper. Other details of the SIDM implementation are given by \cite{Fischer_2026b}.

Finally, we point out that our implementation of unequal-mass scattering is publicly available. This took place as part of the \textsc{OpenGadget3} code release \citep{Dolag_2026, Fischer_2026b}.\footnote{\url{https://gitlab.lrz.de/AstroCodes/OpenGadget3}}

\section{Thermalisation test problem} \label{sec:tests}
In this section, we detail the test problem we employ to ensure that our implementation of two DM species interactions works as expected. In the following, we give the setup of the test problem, the simulation results, and the corresponding analytical solution.

\begin{figure*}
\noindent \begin{centering}
\includegraphics[width=0.9\textwidth]{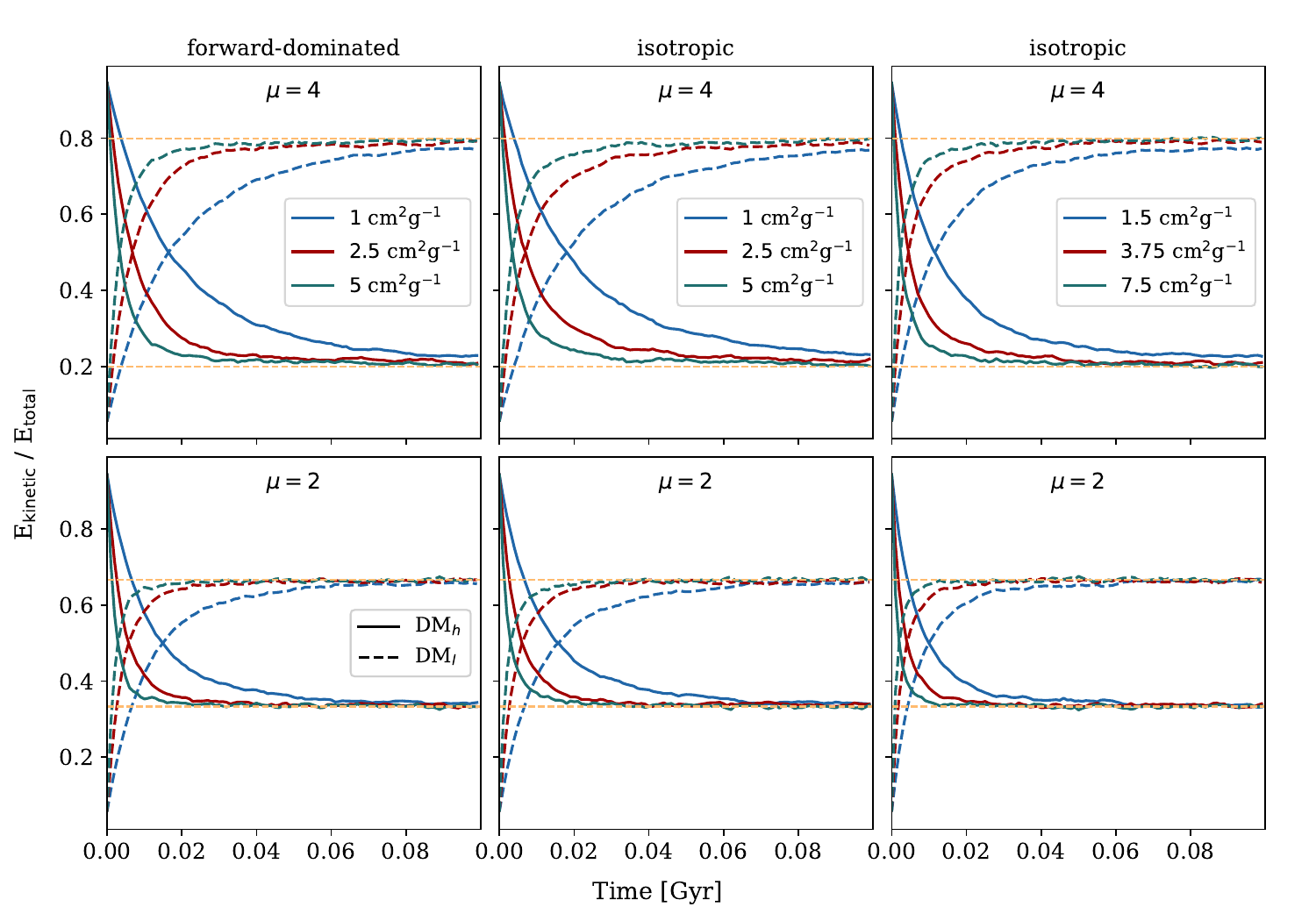}
\par\end{centering}
\caption{\label{fig: tp1}Kinetic energy of $\mathrm{DM_{h}}$ and $\mathrm{DM_{l}}$ particle species as a function of time for different DM models. 
We show models with two different mass ratios, $\mu = 4$ (\textit{top row}) and $\mu=2$ (\textit{bottom row}), and three different momentum transfer cross-sections~(Eq.~\ref{eq:momentum_transfer_cross_section}) in forward-dominated (\textit{left}), isotropic (\textit{middle}), and isotropic viscosity-matching (\textit{right}) scenarios.
The dashed lines represent $\mathrm{DM_{l}}$ and the solid lines represent $\mathrm{DM_{h}}$. The horizontal dashed lines show the theoretical equilibrium states.  We observe that the higher cross-section evolves to the equilibrium state faster compared to the lower cross-sections.}
\end{figure*} 

\begin{figure}
    \centering
    \includegraphics[width=\columnwidth]{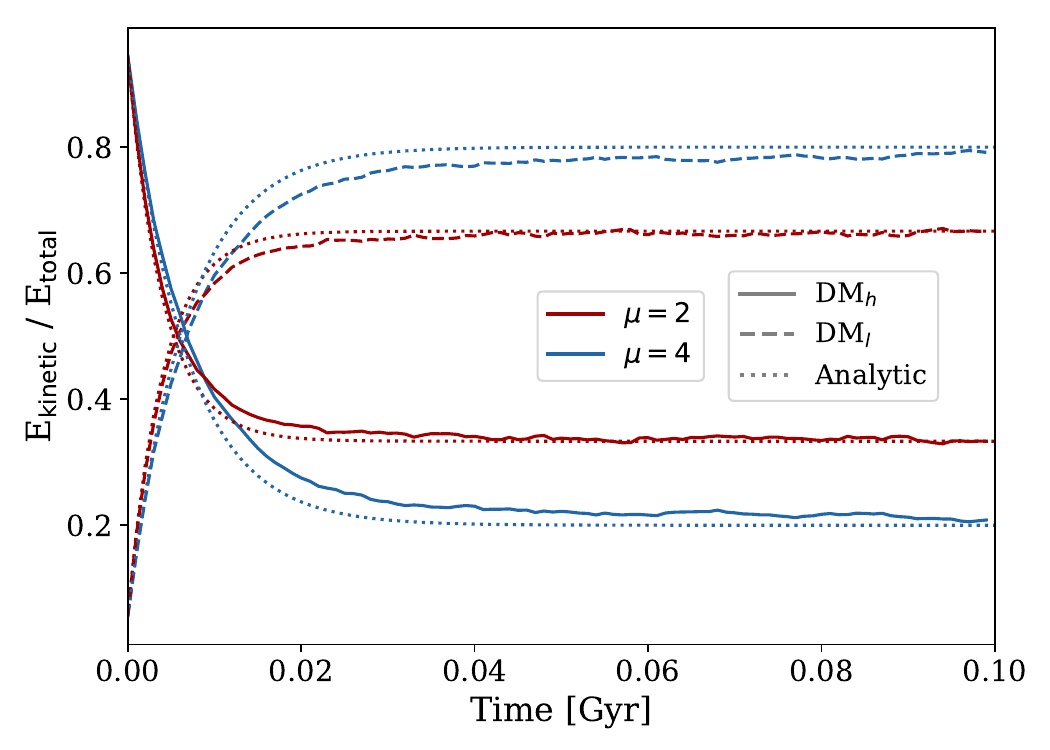}
    \caption{Analytic solution for the fSIDM, $\sigma_\mathrm{T}/m_\mathrm{DM_l} = 2.5$ cm$^2$g$^{-1}$ cross-section scenario for $\mu = 2$ and 4. We show the kinetic energy relative to the total energy for the heavy and light species as a function of time.}
    \label{fig:analytic_tp1}
\end{figure}

Our test relates to the thermalisation problem. We set up a periodic cubic box of length 10~kpc populated with $\mathrm{DM_{h}}$ and $\mathrm{DM_{l}}$ particles. The velocity components for each particle are drawn from a Gaussian distribution with varying one-dimensional velocity dispersion, $\nu$. In this way, we were able to obtain a Maxwell-Boltzmann distribution for the absolute velocities with different temperatures for each species. For the relatively hotter ($\mathrm{DM_{h}}$) species, we employed $\nu = 4 \, \mathrm{km} \, \mathrm{s}^{-1}$ and for the colder species, ($\mathrm{DM_{l}}$) $\nu = 1 \, \mathrm{km} \, \mathrm{s}^{-1}$. The mass of each species is kept constant and equal to 8 $\times$ $10^{13}$ M$_{\odot}$ over all test problem scenarios. To isolate the effect of DM scattering, gravity was excluded and forward-dominated cross-sections (as well as isotropic scatterings) were studied.
Moreover, we used the momentum transfer cross-section (Eq.~\ref{eq:momentum_transfer_cross_section}) and tested the viscosity cross-section (Eq.~\ref{eq:viscosity_cross_section})
for matching different angular dependencies.

Given that the relative velocity of the species is zero, we expected no net momentum transfer between the components over the course of the simulation. In this scenario, the species at higher temperature lose energy via the exchange with the species at lower temperatures, eventually reaching an energy equilibrium stage.

Figure~\ref{fig: tp1} shows the kinetic energy, $E_\mathrm{kinetic}$, for each species evolving with time as a fraction of the total kinetic energy, $E_\mathrm{total}$, given by the sum of the two species. Two scenarios with $\mu=2$ and $\mu=4$ are depicted and the resolution of DM particles in each species is varied. For the $\mu = 4$ scenario, we employ $10^4$ particles and for the $\mu = 2$ scenario, $1.2\times10^4$ particles. Momentum transfer cross-sections of 1, 2.5, and 5 cm$^2$ g$^{-1}$ were simulated for forward-dominated scattering (left column) as well as their isotropic counterparts when matched with the momentum transfer cross-section (middle column) and viscosity cross-section (right column). We find that the higher cross-section scenario evolves towards equilibrium faster compared to the lower cross-sections. We can also note that the momentum transfer cross-section is the comparatively more relevant quantity for matching models of different angular dependences, in contrast to the viscosity cross-section for a single-species model \citep[e.g.][]{Yang_2022D, Sabarish_2024}. This can be seen when comparing the left panels to the middle ones, which corresponds to a matching with the momentum transfer cross-section. In contrast, comparing the left and right panels corresponds to matching the cross-sections based on the viscosity cross-section. The fact that $\sigma_\mathrm{T}$ is the more relevant quantity can also be understood in terms of the exact solution for this problem, which we discuss below.

The exact solution for our test problems can be derived following \cite{Dvorkin_2014} \citep[see also][]{Munoz_2015}. We express the heat exchange between the two species in terms of the time derivative of the energy density $w_\mathrm{l}$ of the lighter species. In the velocity-independent case, it is
\begin{equation}
    \frac{\mathrm{d}w_\mathrm{l}}{\mathrm{d}t} = - 8 \sqrt{\frac{2}{\uppi}} \, \frac{\rho_\mathrm{l} \, \rho_\mathrm{h}}{(1+\mu)^2} \, \frac{\sigma_\mathrm{
    T}}{m_\mathrm{l}} \, \sqrt{\nu^2_\mathrm{h} + \nu^2_\mathrm{l}} \, \left( \nu^2_\mathrm{l} - \mu \, \nu^2_\mathrm{h} \right) \,.
\end{equation}
Here, $\nu^2_\mathrm{h}$ is the velocity dispersion of the heavier species and $\nu^2_\mathrm{l}$ is the velocity dispersion of the lighter species, respectively. We also compare the simulation results to the exact solution. In our case, with a velocity-independent cross-section, the analytic description can be simplified to
\begin{equation} \label{eq:heat_conduct_exact_solution}
    E_\mathrm{{DM}_{l}}(t) = E_\mathrm{{DM}_{l}, eq} + (E_\mathrm{{DM}_{l}, ini.} - E_\mathrm{{DM}_{l}, eq}) \, e^{- \kappa t} \,.
\end{equation}
Here, $E_\mathrm{{DM}_{l}, ini}$ is the initial energy of the lighter DM and $E_\mathrm{{DM}_{l}, eq} = (E_\mathrm{tot} \, \mu) / (1+\mu)$, with $E_\mathrm{tot}$ being the total energy (i.e.\ the sum of $\mathrm{{DM}_{l}}$ and $\mathrm{{DM}_{h}}$). The speed at which the heat transfer happens is set by
\begin{equation} \label{eq:heat_conduct_kappa}
    \kappa = \frac{8}{\sqrt{\uppi}} \, \left(\frac{2}{3}\right)^{3/2} \frac{\sqrt{w_\mathrm{tot} \, \rho_\mathrm{tot}}}{1+\mu} \, \frac{\sigma_\mathrm{T}}{m_\mathrm{{DM}_{l}}} \,.
\end{equation}
The total energy density, $w_\mathrm{tot} = w_\mathrm{{DM}_{l}} + w_\mathrm{{DM}_{h}}$, is given by the sum of the energy density (i.e.\ energy per volume) of the two species. Analogously, the total matter density is  $\rho_\mathrm{tot} = \rho_\mathrm{{DM}_l} + \rho_\mathrm{{DM}_h}$.
We note that the equations above are only valid when $\rho_\mathrm{{DM}_l} = \rho_\mathrm{{DM}_h}$.

\begin{table*}
    \caption{Simulation parameters.}
    \centering
    \renewcommand{\arraystretch}{1.2}
    \begin{tabular}{c c c c c c c}
        \hline
        $f$  & $\mu$  & ${\sigma_\mathrm{T}}/{m_\mathrm{DM_{l}}}$ [cm$^2$ g$^{-1}$] & $N_\mathrm{DM_h}$ [$\times 10^6$] & $N_\mathrm{DM_l}$ [$\times 10^6$] & $m_\mathrm{n,h}$  [$10^4 \, \mathrm{M_{\odot}}$] & $m_\mathrm{n,l}$  [$10^4 \, \mathrm{M_{\odot}}$] \\
        \hline
        0.05  & 1.5  & 1333.33  & 2.085366 & 0.164634 & 7.11 & 4.74 \\
             & 4.0  & 500.00   & 1.858696 & 0.391304 & 7.98 & 1.99 \\
        \hline
        0.125 & 4.0  & 200.00   & 1.431818 & 0.818182 & 9.54 & 2.38 \\
        \hline
        0.2   & 1.5  & 333.33   & 1.636364 & 0.613636 & 7.63 & 5.09 \\
             & 4.0  & 125.00   & 1.125000 & 1.125000 & 11.09 & 2.77 \\
        \hline
        0.5   & 1.5  & 133.33   & 0.900000 & 1.350000 & 8.67 & 5.78 \\
             & 4.0  & 50.00    & 0.450000 & 1.800000 & 17.34 & 4.34 \\
        \hline
        1.0 (Single Species)   & —    & 100        & — & 2.250000       & — & 6.94   \\
        \hline
    \end{tabular}
    
    \tablefoot{The table lists all the parameters for our simulation scenarios. The first column contains the mass fraction, $f$, the second column gives the mass ratio, $\mu$, of DM particles, and the third column displays the momentum transfer cross-section of self-interactions (Eq.~\ref{eq:momentum_transfer_cross_section}). Columns 4 and 5 specify the number of heavier and lighter numerical particles, respectively. Finally, columns 6 and 7 contain the numerical masses of heavier and lighter particles.}
    \label{tab:extended_values}
\end{table*}

Figure~\ref{fig:analytic_tp1} shows the analytic solution for the fSIDM, $\sigma_\mathrm{T}/m_\mathrm{DM_l} = 2.5$ cm$^2$ g$^{-1}$ cross-section scenarios with $\mu = 2$ and $4$. The analytical solution assumes that each component follows a Maxwell-Boltzmann distribution, which does not hold in the simulations beyond the initial and final stages. However, in the beginning, we see that the simulated gradient, $\mathrm{d}E_\mathrm{kinetic}/\mathrm{d}t$, matches the analytical expectation very well. Moreover, our simulations correctly reproduce the equilibrium state at late times. Therefore, we assume that the simulation code works properly.

\section{Isolated halo}\label{sec:isolated_halo}
In this section, we study how the evolution of an isolated DM halo is altered by the interactions of our two-species model. We first describe the simulation set-up in Sect.~\ref{sec:simulation_setup}, followed by the simulation results in Sect.~\ref{sec:halo_evolution}.
Here, we describe and qualitatively explain the evolution of the DM halo with the two-species model considered in this work.
Finally, in Sect.~\ref{sec:halo_projection}, we show the enclosed mass and the density slope of the halo in projection as commonly done for studies of strong gravitational lensing.

\subsection{Simulation setup}\label{sec:simulation_setup}

Our simulation setup consists of a single isolated halo comprised of two DM species, where the population and masses of DM species are governed by two parameters: mass ratio, $\mu$, and mass fraction, $f$. 
The initial conditions (IC) for the DM halo are generated using the publicly available code \textsc{SpherIC} \citep{Rocha_2013b} by Miguel Rocha, which we modified to account for unequal particle masses. The velocities of particles were assigned using the Eddington inversion method, assuming an isotropic velocity distribution. Both DM species are initially drawn from the same velocity distribution function. We tried to replicate the setup used by \cite{Zhong_2023}, also studied by \cite{Fischer_2024a}. The DM halo follows a Navarro–Frenk–White (NFW) profile \citep{Navarro_1997},
\begin{equation}
\rho_{\rm NFW}(r)=\frac{\rho_0}{(r/r_\mathrm{s})(1+r/r_\mathrm{s})^2} .
\end{equation}
It is described by the following parameters: scale radius, $r_\mathrm{s} = 9.1 \, \mathrm{kpc}$, and scale density, ${\rho}_0$ = 6.9 $\times$ $10^6$ M$_{\odot}$ kpc$^{-3}$. This corresponds to a mass of $M_{200} \approx 10^{11} \, \mathrm{M_\odot}$ and a concentration parameter of $c_{200} \approx 10.5$.
The number of numerical particles is kept constant at $2.25\times10^6$ particles for all simulations, while the number of particles of each species and their masses is varied for different values of $\mu$ and $f$ as listed in Table~\ref{tab:extended_values} to keep the mass of the halo constant. When sampling the initial conditions, we choose a cut-off radius of $15 \times r_\mathrm{s}$, after which the halo follows an exponentially decaying profile outward to avoid a sharp truncation. The ICs generated from \textsc{SpherIC} are modified for our simulations by randomly choosing particles and altering their masses to obtain a DM halo with the desired number of light and heavy DM particles. Moreover, we set the gravitational softening length to $\epsilon = 0.13$ kpc.

In Appendix~\ref{sec:stability_test}, we describe how we tested the stability of the ICs for collisionless DM. The evolution of the halo reveals a density profile that stays constant over time for the simulated period of 10~Gyr. No strong signs of mass segregation are visible in the CDM simulations due to the absence of energy transfer from DM scattering. Moreover, the chosen numerical resolution is high enough such that there is no significant artificial gravitational relaxation.

Table~\ref{tab:extended_values} lists all the SIDM scenarios we simulate in this study and the range of parameters considered. The cross-section for each scenario is adjusted such that $f \, \mu \, \sigma_\mathrm{T}/m_\mathrm{DM_{l}} = 100 \, \mathrm{cm}^2 \, \mathrm{g}^{-1}$ holds true.

\subsection{Halo evolution}\label{sec:halo_evolution}

Our study is focussed on the evolution of DM halos, starting from a configuration where both DM species follow the same distribution. We made the assumption that the extent of dynamical mass segregation is minimal during the formation of the halo at high redshift. This is because the strength of interactions at early times depends on the velocity dependence of the cross-section and since we have a constant cross-section, SIDM becomes relevant at rather late cosmic times. Our simulations begin with no observable mass segregation initially and evolve for a time period of no more than 20 Gyr. We only used simulation data with a relative error in energy conservation of less than $0.5 \%$ (see Appendix~\ref{sec:energy_cons}).

In the following, we examine various aspects of the halo evolution with our two-species model and compare it to the common single-species SIDM case. This allows us to gain a qualitative understanding of how the halo evolves under the interaction of the two DM species.

\subsubsection{Density profile}

\begin{figure}
    \centering
    \includegraphics[width=\columnwidth]{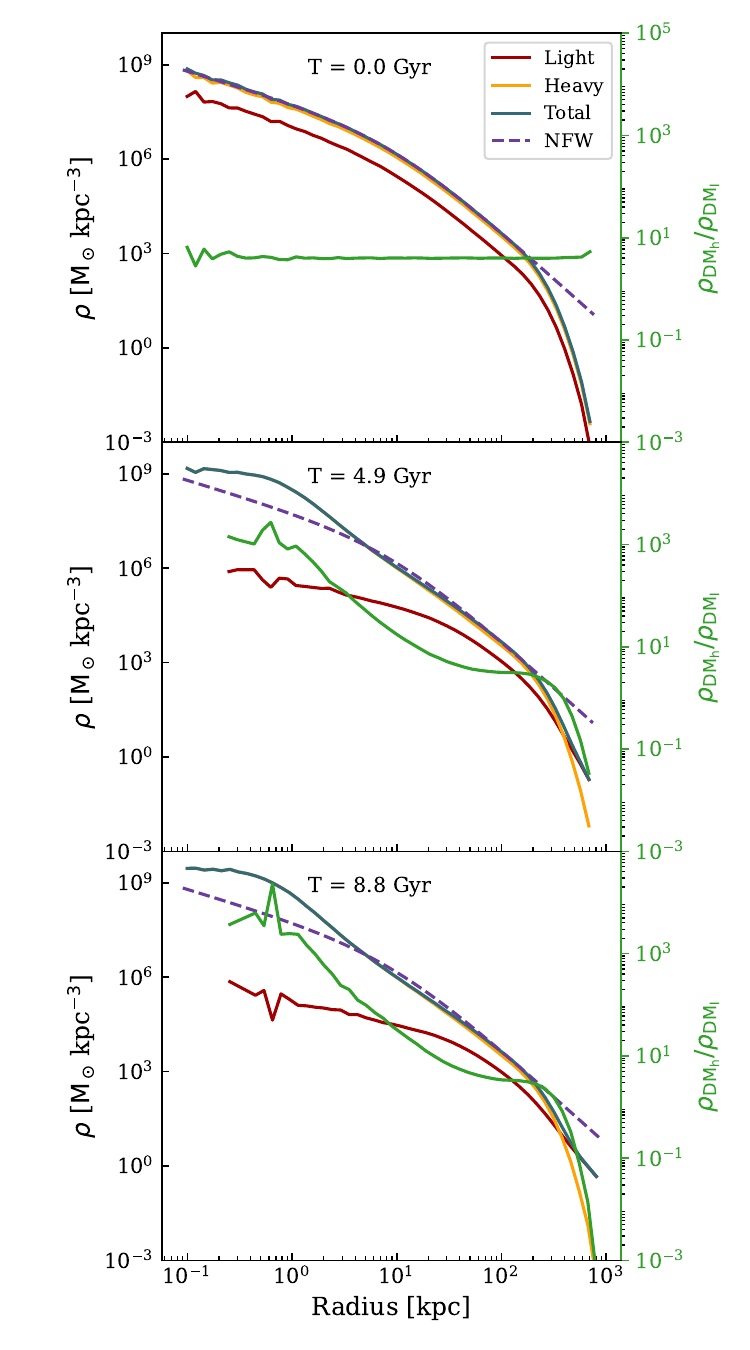}
    \caption{Evolution of the density profile. The density as a function of radius is displayed for the total mass and the two species individually.
    The results are for the simulation with $f=0.2$ and $\mu=4$.
    \textit{Top}: Initial density profile. The green curve indicates the density ratio between the species. It shows an equal distribution of particles initially. \textit{Middle}: Effect of mass segregation as the total density rises above the NFW curve in the inner region. \textit{Bottom}: State close to the final stable density profile, which avoids the gravothermal collapse phase and has a significantly higher central density region.}
    \label{fig:density_evolution}
\end{figure}

We begin with the density profiles to discuss the stages of evolution for the scenario $f=0.2$ and $\mu=4$, which are qualitatively similar to the other models we study in this work. Figure~\ref{fig:density_evolution} shows the evolution of density for both species and the total density, beginning from $T=0$ Gyr (top panel). Prior to this moment, we assumed the evolution of the halo to be influenced mainly by gravity alone. The red curve shows that it follows an NFW profile with a cut-off length after which it displays an exponentially decaying profile.\footnote{This artificial truncation is introduced when sampling the ICs to deal with the non-finite mass of an NFW halo.} The green curve shows the density ratio of $\mathrm{DM_{h}}$ and $\mathrm{DM_{l}}$, affirming that they are evenly distributed throughout the halo.

Dynamical mass segregation begins after $T=0$, as scatterings between the $\mathrm{DM_{h}}$ and $\mathrm{DM_{l}}$ particles lead to an exchange of energy. As explained in Sect.~\ref{sec:model}, the $\mathrm{DM_{l}}$ particles gain energy from $\mathrm{DM_{h}}$ particles and carry it to outer regions. The $\mathrm{DM_{h}}$ particles lose energy and move to the inner orbits. This effect changes the density profile of the DM halo by making the inner regions denser compared to the initial profile. The middle panel of Fig.~\ref{fig:density_evolution} shows the evolved density profile at 5 Gyr. The total density overshoots the initial NFW profile in the inner regions, and a comparatively denser core has formed. The density ratio of both species (green curve) shows an effective movement of heavier particles towards the inner regions and lighter particles towards the halo outskirts. The ratio of densities between the species rises in the inner regions, confirming the movement of particles in the specified directions.
Over time, some particles gain sufficient energy to leave the gravitationally self-bound DM halo. However, the unbound fraction remains below 0.5\% of the total number of particles throughout the simulation.

The bottom panel of Fig.~\ref{fig:density_evolution} shows the almost final stage of evolution of the DM halo. The density observed in the inner regions is significantly larger due to mass segregation. The density ratio of both species also indicates increased concentration of $\mathrm{DM_{h}}$ particles in the inner regions. It is important to note that this is expected to be close to the final state of the DM halo with the observed inner density, which starts to plateau during this time. At this moment, very few $\mathrm{DM_{l}}$ particles remain in the inner regions of the halo that can carry energy to outer orbits, and mass segregation effectively ceases. The simulation does not experience a gravothermal catastrophe, as known from SIDM simulations with identical particles.

\subsubsection{Evolution of the central density}

\begin{figure*}
    \centering
    \includegraphics[width=\columnwidth]{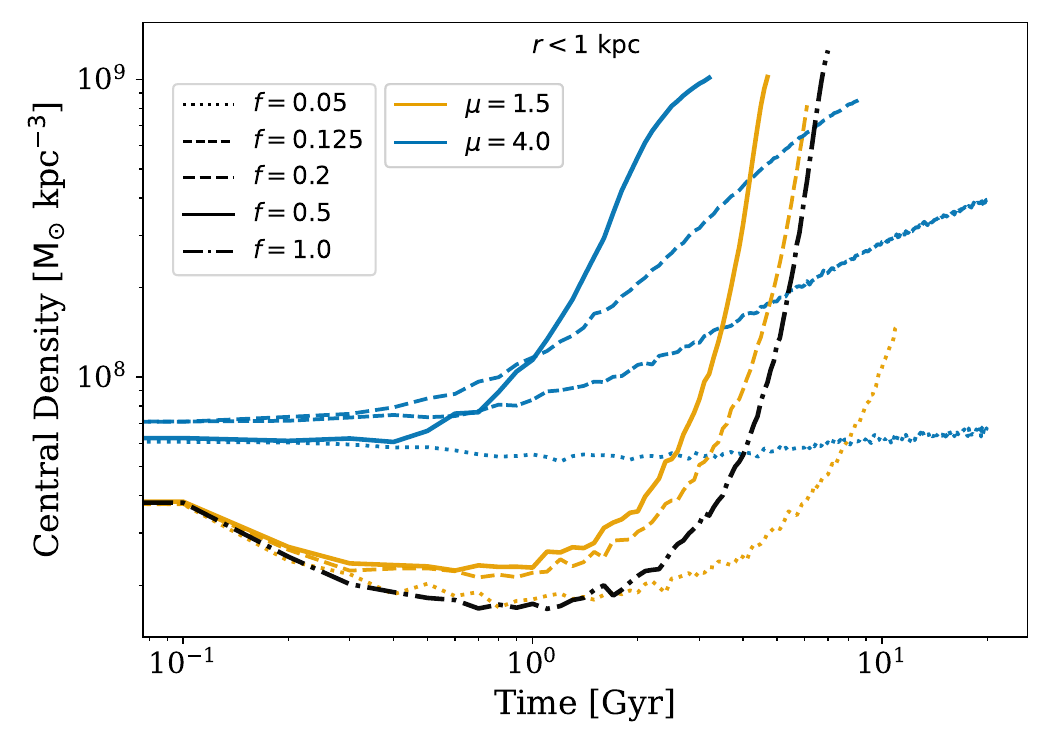}
    \includegraphics[width=\columnwidth]{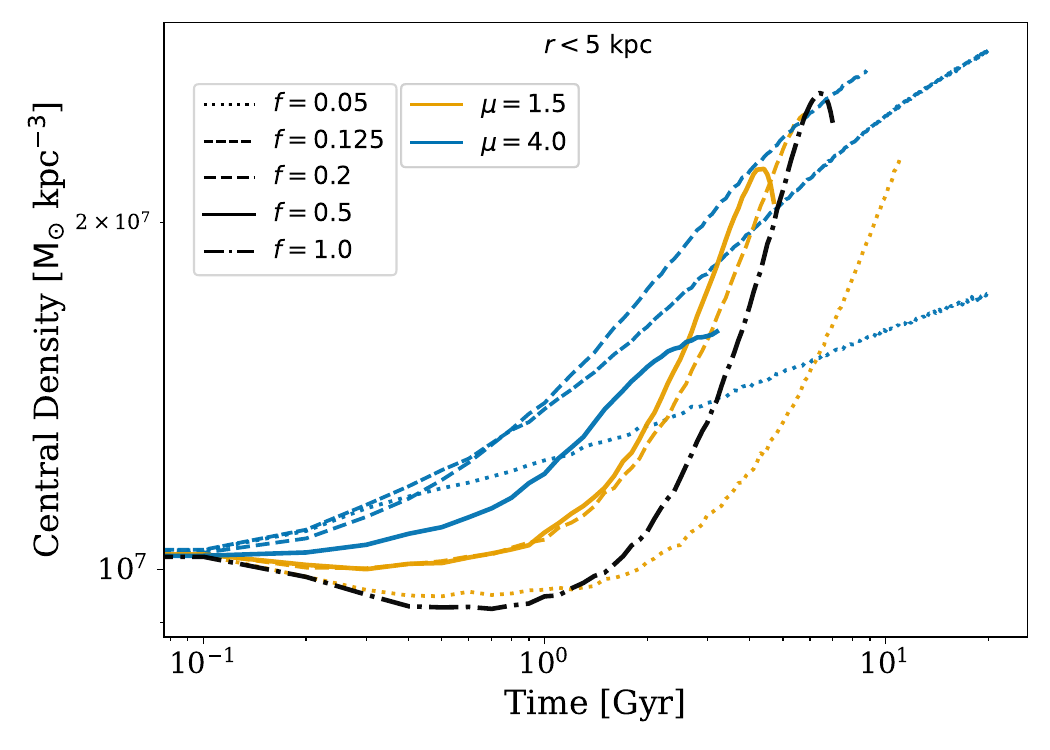}
    \caption{Central density as a function of time. \textit{Left}: Average density within 1 kpc. \textit{Right}: Average density within 5 kpc. Initially, the simulations have the same central density and then move to lower densities, showing a core expansion phase, and as mass segregation continues, they reach higher densities that plateau after a certain time. For most of the two-species models, we do not observe a sharp rise in central density at later stages as observed in the single-species SIDM scenario (black) during the gravothermal collapse phase.}
    \label{fig:central_density}
\end{figure*}

Next, we studied how the evolution of the DM halo depends on the model parameters; namely the mass ratio, $\mu$, and the mass fraction, $f$. For this, we focussed on how the central DM density of the halo evolves with time, as is commonly done in studies of SIDM halos.

In Fig.~\ref{fig:central_density}, we study the average central density of the halo for our simulations inside a radius of 1 kpc (left panel) and 5 kpc (right panel). The lower density initially depicts the core expansion phase in SIDM simulations. With time, the density increases to the maximum value and plateaus, avoiding a sharp increase generally observed in gravothermal collapse scenarios with single DM species also simulated and depicted here as the black curve.
For the different simulations in our study, the resulting halo characteristics depend on the chosen parameters.
In particular, how pronounced the core expansion phase is (i.e.\ how small the minimum density reached during the evolution is) clearly depends on the mass ratio $\mu$ of the DM particles. The more unequal the particle masses, the smaller the density core that forms.
In Sect.~\ref{sec:Knudsen_number}, we discuss this dependence further and also explain the role of the mass fraction for the core formation.
Moreover, the central density that the halo approaches at late stages depends on the mass fraction, $f$. Although we were not able to fully simulate this stage for all models, due to numerical limitations, it is visible that the central density at the late stages increases with $f$ (see the left panel).

The right panel for the average density within 5~kpc shows a decrease in the density at late times for a few models. This is a feature known for single-species models. Right from the beginning of the simulation, the halo loses mass at large radii to even larger radii. With time, the region that loses mass extends to smaller and smaller radii \citep{Fischer_2025b}. We observed a similar behaviour for a few of the two-species models. The details of this transition from mass growth to mass loss within a fixed radius depend on the model parameters $f$ and $\mu$. For example, the maximum average density within 5~kpc decreases with an increasing value for $f$. However, this may not be true across the full range of possible values for $f$.
Overall, the interplay between the two-species gives rise to a non-trivial halo evolution.

\subsubsection{Density growth during the late stages of halo evolution}

\begin{figure}
    \centering
    \includegraphics[width=\columnwidth]{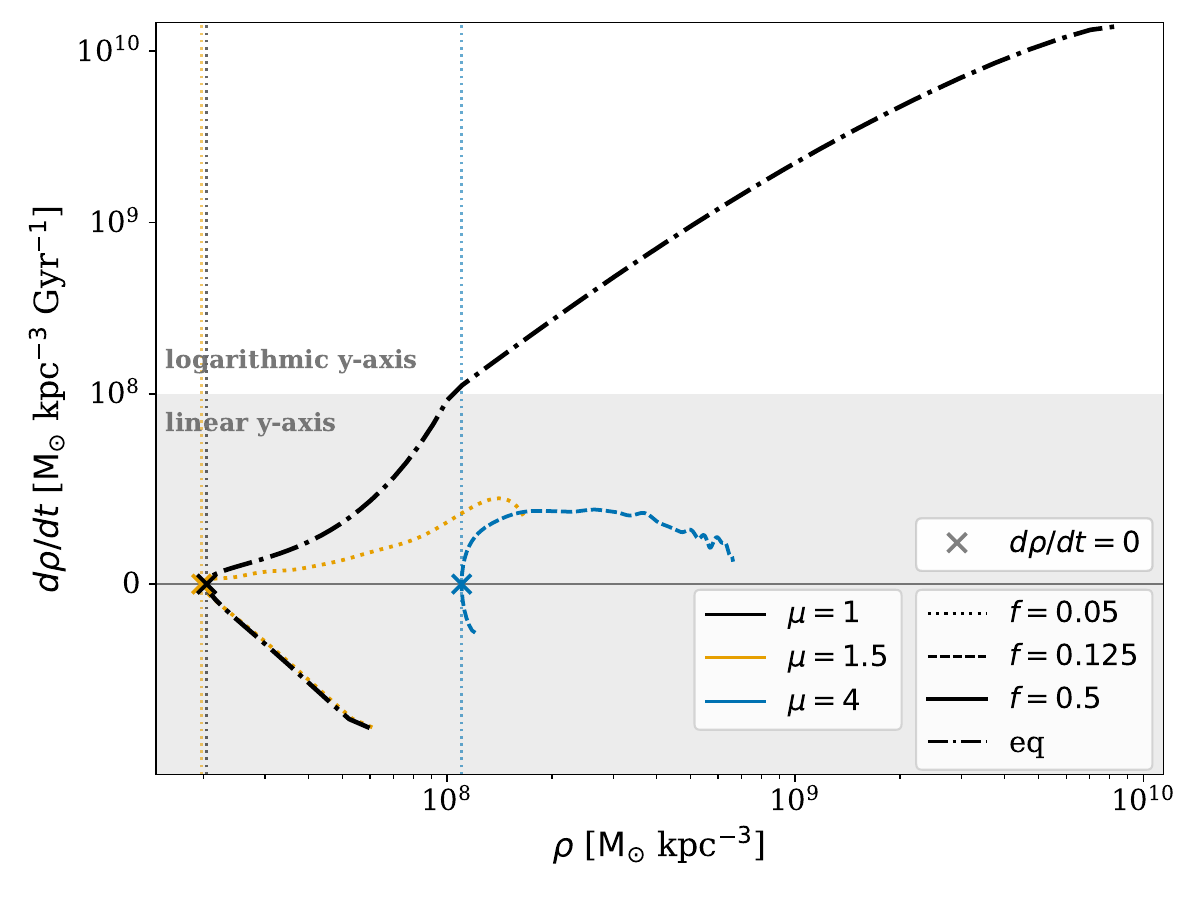}
    \caption{Time derivative of the central density as a function of the central density. We show the halo evolution for the single-species scenario and two two-species models in terms of the growth of the central density at different evolution stages characterised by the central density. The regime of low growth rates is displayed using a linear y-axis (gray area) whereas for large values we use a logarithmic y-axis. Maximum core formation is reached about the point when the central density is the lowest marked by crosses. Additionally we indicated those densities by vertical lines.}
    \label{fig:drho/dt_vs_cen_den}
\end{figure}

As the evolution of the central density at late stages is a remarkable feature of the two-species model, we took a closer look at it. Therefore, we show the time derivative of the central density as a function of the central density in Fig.~\ref{fig:drho/dt_vs_cen_den}.

Here, we use the density at the very centre of the halo. It is determined by fitting a King profile \citep{King_1962} to the innermost 2~kpc of the halo, 
\begin{equation} \label{eq:king_profile}
    \rho(r) = \rho_0 \left[ 1 + \left( \frac{r}{r_\mathrm{c}} \right)^2 \right]^{-3/2}
\end{equation}
The King profile depends on the parameters, $\rho_0$, which gives the density at the centre of the halo, and $r_\mathrm{c}$, which specifies the spatial extent of the central core. 
For the fit, we maximised a likelihood based on Poisson statistics analogously to the description in Sect.~4 by \cite{Fischer_2021a}. The fitted value for $\rho_\mathrm{c}$ is then used for Fig.~\ref{fig:drho/dt_vs_cen_den}, after applying a Gaussian filter to reduce noise.

The figure shows that initially, the central density decreases during the core formation phase and subsequently increases during the collapse phase. This is a common behaviour for a single-species SIDM model and our two-species cases.

However, at late times, we find a qualitative difference in the behaviour of the single-species scenario and the two-species models. For the central density of the two-species scenarios at late time, we observe a decaying gradient, $\mathrm{d}\rho/\mathrm{d}t$. It indicates a decelerating density increase with time and hints toward the asymptotic approach to a stable dense halo. This behaviour is not present in the single-species case.

In our two-species model, the final density depends on how much energy can be transported outward. The mass fraction, $f$ of the light species determines how much mass is available to transport energy outward and thus is critical in setting the value of the final density. The mass ratio, $\mu$, determines how efficient this transport per scattering is. Hence, both these parameters govern the evolution towards the final central density approached by the halo.

\subsubsection{Knudsen number}
\label{sec:Knudsen_number}

\begin{figure*}
    \centering
    \includegraphics[width=0.95\textwidth]{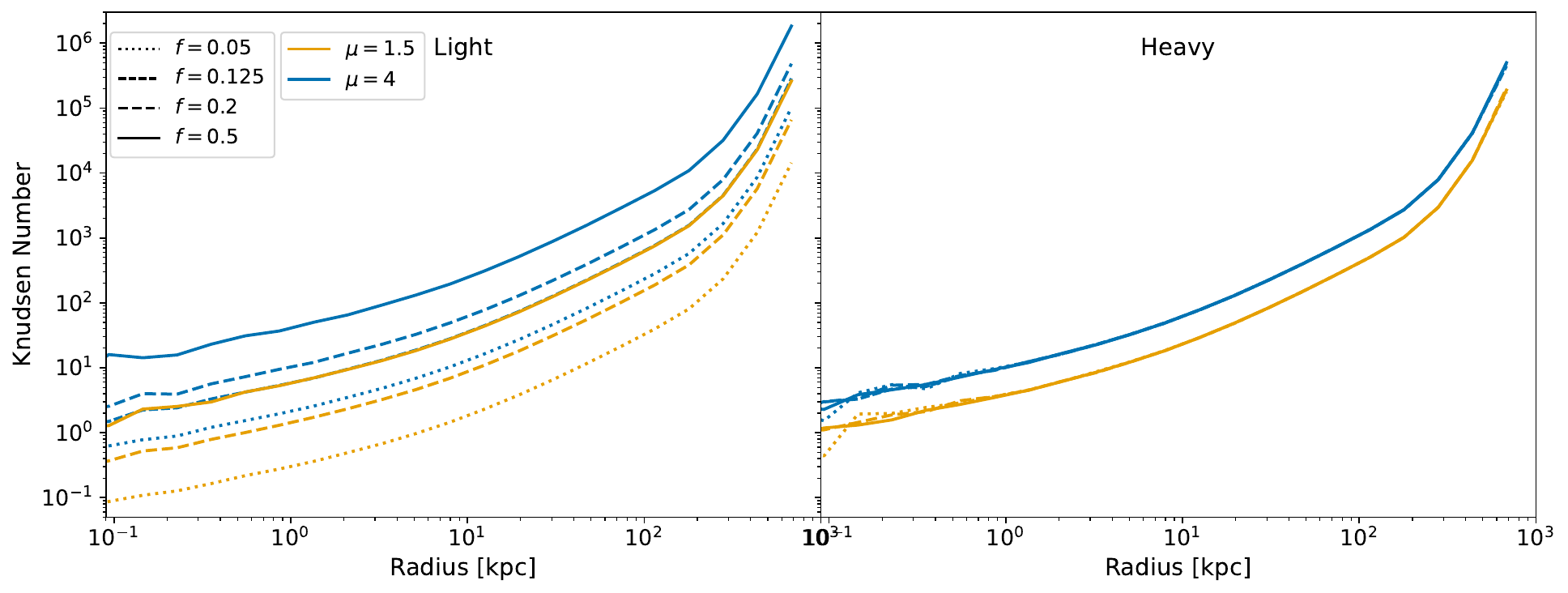}
    \caption{Knudsen number as a function of radius for both light and heavy species. We show the Knudsen number (Eqs.~\ref{eq:Knudsen_light} and~\ref{eq:Knudsen_heavy}) based on our ICs for the different scenarios we studied. \textit{Left}: Values for the light species. \textit{Right}: Values for the heavy species. The lighter particles in lower mass fraction scenarios lie in the SMFP regime, and higher mass fractions tend towards the LMFP regime.}
    \label{fig:Knudsen_no}
\end{figure*}

To gain further insights into the halo evolution, we study the Knudsen number for both species.
It compares how much the trajectory of a DM particle is affected by DM scattering relative to gravity. 
The Knudsen number is defined as 
\begin{equation}
    {Kn} = \frac{\lambda}{H} \, ,
    \label{eq:knudsen_number}
\end{equation}
where $\lambda$ is the mean free path of scattering and $H$ is the gravitational scale height, given by the Jeans length, $H = \sqrt{\nu^2 / (4 \uppi \, \mathrm{G} \, \rho)}$. For $\mathrm{DM_l}$, it is given as
\begin{equation}
    {Kn}_\mathrm{l} = \frac{\mu}{\rho_\mathrm{h} \, \nu_l} \sqrt{4 \uppi \, \mathrm{G} \, \rho} \left(\frac{\sigma_\mathrm{T}}{m_\mathrm{l}}\right)^{-1} \, ,
    \label{eq:Knudsen_light}
\end{equation}
where $\rho_\mathrm{h}$ is the matter density of $\mathrm{DM_h}$, $\rho$ is the total matter density of both species, $\nu$ is the 1D velocity dispersion, and G is the gravitational constant. All quantities are computed from spherically averaged radial profiles: the velocity dispersion and density values both are evaluated in logarithmic shells. Similarly, for $\mathrm{DM_h}$ it is defined as 
\begin{equation}
    {Kn}_\mathrm{h} = \frac{1}{\rho_\mathrm{l} \,\nu_h} \sqrt{4 \uppi \, \mathrm{G} \, \rho} \left(\frac{\sigma_\mathrm{T}}{m_\mathrm{l}}\right)^{-1} \, ,
    \label{eq:Knudsen_heavy}
\end{equation}
where $\rho_\mathrm{l}$ is the matter density of $\mathrm{DM_l}$. If the mean free path for a particular species is smaller than the gravitational scale height, resulting in ${Kn} \ll 1$, the system is in the short-mean-free-path (SMFP) regime. In this regime, a typical particle undergoes multiple scattering events before crossing the system once, and its trajectory is dominated by the DM self-interactions rather than gravitational interactions. Conversely, if ${Kn} \gg 1$, the species is in the long-mean-free-path (LMFP) regime. In this case, a DM particle may traverse the system multiple times, experiencing little energy exchange via scattering, and its trajectory is dominated by the gravitational potential rather than self-interactions. 

Figure~\ref{fig:Knudsen_no} shows the Knudsen number for both light and heavy particles computed from the initial NFW profile. In most scenarios, we observe that the lighter and heavier species both exist in the LMFP region, for which we also observe the formation of a density core for sufficiently small values of $\mu$ as shown in Fig.~\ref{fig:central_density}. The core expansion happens due to heat flowing inward into the central cool region of the NFW halo. This heat comes from intermediate radii with a larger velocity dispersion. For our two-species model, this is more complex than in a typical single-species model. To make heat flow inward, the two species must interact, with energy transfer between heavier particles effectively mediated by the lighter species. However, in every interaction, the mass segregation plays a role and tends to make the light particles move outward, which effectively acts against the heat inflow. As a consequence, core expansion is stronger suppressed in models with a larger mass ratio $\mu$. However, the picture changes for the $f = 0.05$ case with $\mu = 1.5$, where we observe that the lighter particles have a very low Knudsen number and lie in the SMFP regime, in contrast to the heavier particles. These lighter particles effectively cannot travel much through the halo due to high energy exchange and mediate larger heat transfer inward, before segregating to larger radii. This, in turn, allows for a more pronounced core formation phase compared to the LMFP regime with larger core sizes. Effectively, we observe a trend of the maximum core size depending on the value of $\mu$ only, except for very small $f$ values, the lighter particles move to the SMFP regime, and we see a longer core formation phase with a stronger decrease in central density (see Fig.~\ref{fig:central_density}).
The dependence on $f$ is also connected to our choice for the cross-section (see Sect.~\ref{sec:simulation_setup}).

In summary, these results demonstrate that the Knudsen number has a distinct impact on halo evolution.  Next, it is crucial to determine the maximum size of the core that a halo can host.

\subsubsection{Summary of halo evolution}
Overall, the halo evolution proceeds in three broad stages.
Initially, the system undergoes a mild core-expansion phase, whose strength depends strongly on the mass ratio of the two species.
This is followed by a phase of dynamical mass segregation, during which the heavier particles migrate inward and the lighter particles carry energy to larger radii, increasing the central density.
At late times, the evolution slows down as the inner region becomes depleted of lighter particles, such that the density approaches a relatively high finite value instead of entering a runaway gravothermal collapse.

\subsection{Halo in projection}
\label{sec:halo_projection}

To compute quantities that are closer to the observations, we studied the projected logarithmic density profile slope and the projected enclosed mass. These quantities are relevant for the strong gravitational lensing analysis of strong lens perturbers \citep[e.g.][]{Tajalli_2025, Li_2025}.
Models of SIDM are known to produce systems with an increased enclosed mass at their centre and steeper density slopes thanks to the gravothermal collapse. This makes SIDM an interesting candidate for explaining observations.
In our two-species model, the late evolution of the halo is qualitatively different compared to a single-species SIDM model, as we show in Sect.~\ref{sec:halo_evolution}. 
Here, we investigate how this affects its prospects in explaining strong lens perturbers.

\begin{figure*}
    \centering
    \includegraphics[width=\columnwidth]{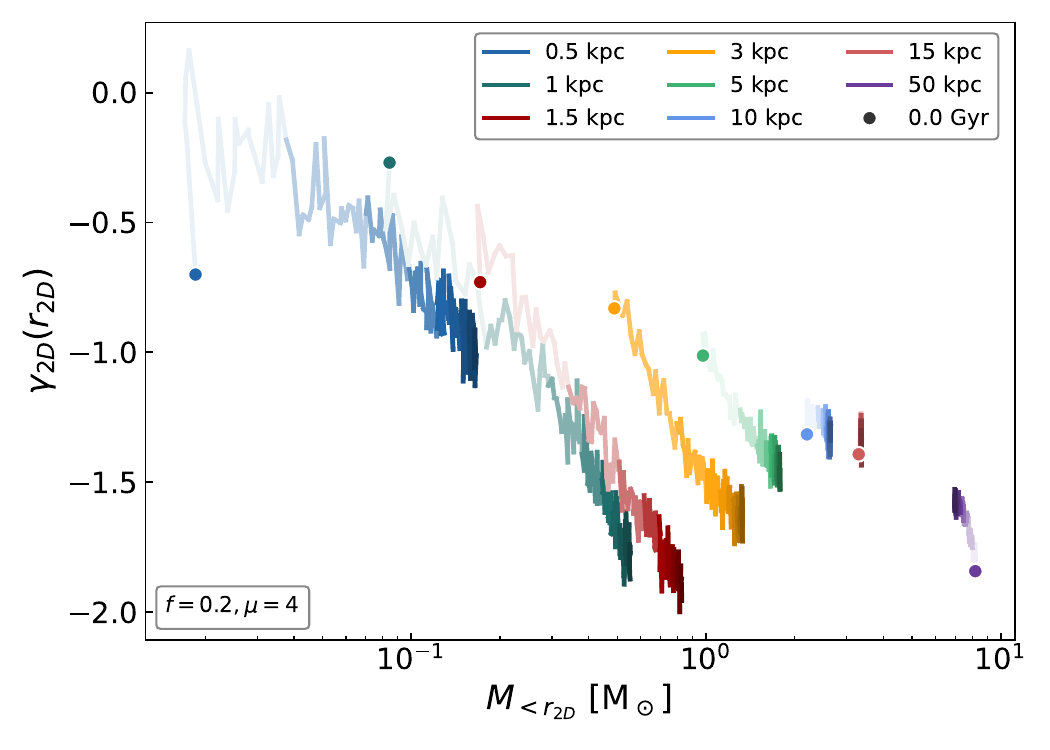}
    \includegraphics[width=\columnwidth]{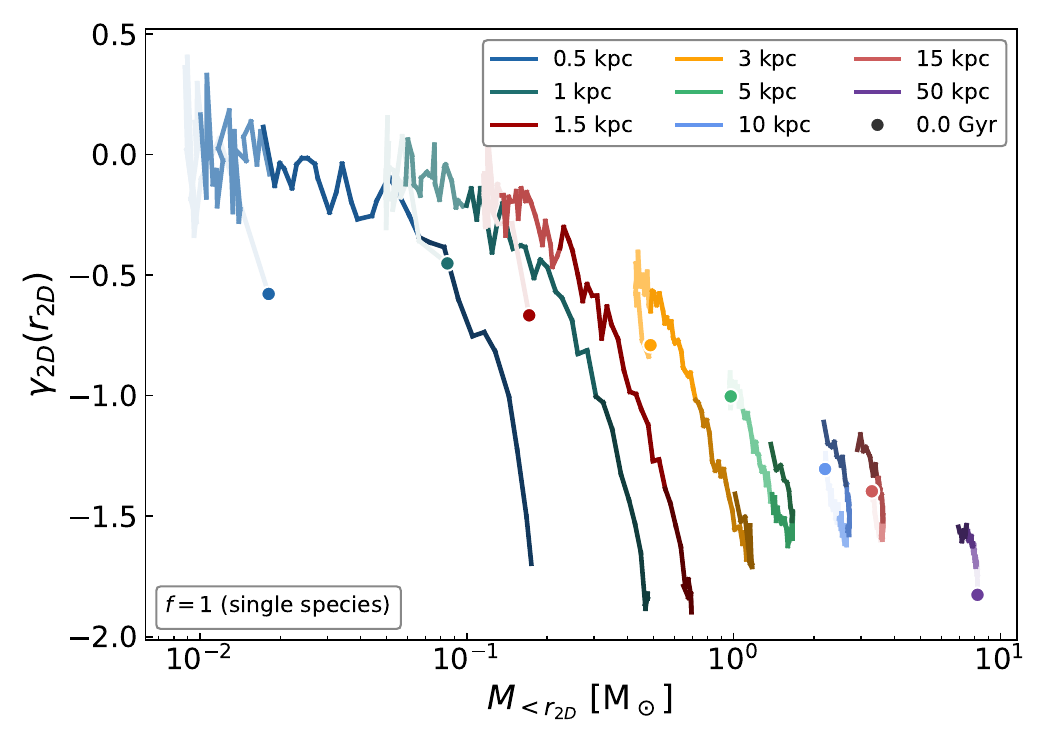}
    \caption{Projected logarithmic density profile slope as a function of projected enclosed mass. \textit{Left}: Scenario with $f=0.2$ and $\mu=4$ (the same simulation as in Fig.~\ref{fig:density_evolution}). \textit{Right}: Single species case. We show how the projected logarithmic density profile slope, $\gamma_\mathrm{2D}$, and the projected enclosed mass, $M_{<r_\mathrm{2D}}$, evolve with time. For several radii $r_\mathrm{2D}$ as indicated in the legend, $\gamma_\mathrm{2D}$ and $M_{<r_\mathrm{2D}}$ are displayed. The larger circles mark the value at the beginning of the simulation, i.e.\ for an NFW profile. Moreover, the colour gradient changes shade at 2 Gyr intervals. }
    \label{fig:slope_vs_mass}
\end{figure*}

In Fig.~\ref{fig:slope_vs_mass}, we show the projected logarithmic density profile slope, $\gamma_\mathrm{2D}$, as a function of the projected enclosed mass, $M_{<r_\mathrm{2D}}$, for our simulation with $f=0.2$ and $\mu=4$ (left panel) and the single-species case (right panel). In the left panel, it is visible how, for all shown radii, except the outermost one, the enclosed projected mass is increasing with time. For radii that are sufficiently small, the projected logarithmic density profile slope becomes steeper as well. Only at the smallest radius does the halo initially undergo a small core formation with a flatter density slope, which later becomes steeper compared to the initial NFW profile.

In the late stages, the evolution of the system slows down and we see that the values $\gamma_\mathrm{2D}$ and $M_{<r_\mathrm{2D}}$ do not change as rapidly. This is illustrated by the different colour shades in Fig.~\ref{fig:slope_vs_mass}, which cover equal time spans. This is in line with the system approaching a stable solution and makes it very promising in the context of explaining strong lens perturbers.
In contrast, for the single-species model with elastic scattering shown in the right panel, the enclosed mass starts to decrease after some time, while the density slope also becomes flatter again~\citep[see also][]{Fischer_2025b}. This is only visible for large radii; however, it is also expected to occur for the smaller radii as well at times later than the ones we have simulated. The increased enclosed mass and the steeper density slope that we find for our two-species model could be promising for explaining surprisingly compact halos~\citep[e.g.][]{Li_2025}. Importantly, in the two-species case, it should be more likely to observe a system with a fairly high enclosed mass and steep density slope, as it is not a transitional phase as for the single-species models, but presumably a stable state. As a consequence, the abundance of fairly compact DM objects could help in discriminating between the DM models.

\section{Discussion} \label{sec:discussion}
In this section, we discuss the shortcomings and limitations of our implementation of the two-species SIDM model. Moreover, we elaborate on the physical implications of DM models with unequal masses and non-gravitational interactions and how to constrain them with the help of observations.

\subsection{Technical aspects}
A limiting aspect of our scheme is that it requires the numerical particles to have the same mass ratio as the physical ones. Given that one may want to achieve a specific mass resolution for the particles of the heavier species, the number of particles required to resolve the light particles increases when choosing a model with more unequal masses (i.e.\ when increasing $\mu$). This can render simulations with larger values of $\mu$ prohibitively expensive. However, we have to note that this problem is also coupled to the mass fraction $f$; namely, for models with a small fraction of light particles, it could become simpler to obtain accurate results.

From SIDM $N$-body simulations, it is known to be challenging to model the late collapse phase of DM halos accurately \citep[e.g.][]{Yang_2022D, Zhong_2023, Mace_2024, Palubski_2024}. In particular, \cite{Fischer_2024b} points out different reasons for energy non-conservation and comments on other limitations as well \citep[see also][]{Fischer_2025b}.
These challenges are also present in our simulation, but the problem is less severe, as we do not necessarily expect the central density to keep growing until it collapses into a BH. This is a stage that is not feasible to simulate with state-of-the-art SIDM $N$-body codes. In contrast, in our model, the final inner density sets the requirements for the $N$-body simulation. Depending on the parameters of the model (i.e. the mass ratio and mass fraction), it can be possible to simulate the very late stages of the halo evolution accurately.

\subsection{Physical considerations}
The single-species SIDM models that are commonly investigated make a specific prediction on the evolution of the inner density of a DM halo. The gravothermal evolution proceeds quickly at late stages and eventually leads to DM collapsing into a BH. The halo spends a relatively short time in its evolution at densities that are higher than expected from CDM, but not by many orders of magnitude. Objects with these intermediate densities are expected to be relatively rare given standard gravothermal evolution.

However, these intermediate densities might be required to explain the gravitational strong lensing results \citep[e.g.][]{Li_2025} and observations of stellar stream gaps \citep[e.g.][]{Bonaca_2019}, which point towards fairly compact satellite halos. This is because the mass in the inner region of the halo starts to decrease at late stages of the gravothermal evolution \citep[e.g.][]{Fischer_2025b}. To gravothermally form a very dense core in the inner region of an SIDM halo, enormous amounts of energy need to be transported outward. This outflow of energy happens via the loss of mass. Specifically, the high-velocity tail of the DM particles can move to larger radii or even become unbound, thereby reducing the energy in the central region. However, this mechanism is inefficient in the sense that it comes with a large mass loss compared to the energy that is transported outward. To obtain a rough estimate of how much mass may reside in the inner region shortly before it could collapse into a BH, we can compare the gravitational binding energy of this inner region with the binding energy of the initial halo \citep[see also the work by][in the context of the formation of primordial black holes]{Ralegankar_2025}. The idea behind this is that the energy released by the inner dense core evaporates or destroys the halo. An estimate for the stage at which a BH would form reveals that the corresponding mass may not be close to what would be required to explain the observations that currently motivate SIDM studies of the gravothermal collapse. This could eventually lead to a timing problem for a single-species SIDM model with elastic scattering.

Additional physical processes could allow for higher BH masses than described above. For example, if a BH were to form from a fraction of the SMFP core, it could accrete the remaining mass of the core in a way that does not require transporting as much energy outward as during the collapse without a BH. However, it is not clear whether accretion from the loss cone or Bondi accretion \citep{Bondi_1952} would allow for the formation of a much more massive BH \citep[see also the work by][for more details on the accretion of SIDM]{Sabarish_2025, Feng_2025}. We note that in the spherical symmetric case, the Bondi flow for a non-conducting fluid in the non-relativistic regime is subsonic; hence, this process of accretion could be less efficient than assumed in previous studies \citep[e.g.][]{Pollack_2015}.

Aside from a potential timing problem, the abundance of halos with a specific central density could provide an insightful measure of the mechanism that makes halos more compact. Specifically, if halos with these intermediately high densities are not as rare as would otherwise be expected from the gravothermal evolution, this could point to a different formation mechanism. In such a case, it would be important to study alternative formation mechanisms for these objects, such as the formation by mass segregation arising from a DM model with unequal particle masses. In this work, we have investigated this kind of model, which comes at the cost of two additional parameters, while allowing for more freedom in the late evolution phase. During this phase the central densities can evolve significantly more slowly and reach lower values, effectively avoiding the gravothermal catastrophe. As a consequence, observations of these late stages would allow us to constrain the additional parameters ($f$ and $\mu$) of our model.

For our simulations, we have assumed an NFW halo for the ICs. However, this might not always be valid. However, for a velocity-independent cross-section, self-interactions only become relevant at low redshifts, making an NFW a reasonable assumption. Subsequent studies taking a similar approach to \cite{Yang_2025, Yang_2026} could offer an investigation of such models within the cosmological context.

\section{Conclusions} \label{sec:conclusion}
In this paper, we study a DM model that consists of two species with unequal masses and non-gravitational interactions across the species. It marks the first opportunity for a qualitative understanding of the impact of such a model and its parameters to be achieved by running $N$-body simulations of an isolated halo. We explain the numerical scheme for unequal-mass scattering and test our implementation in the simulation code \textsc{OpenGadget3}. Finally, we study the effect of our two-species DM model on the evolution of an isolated halo. Our main conclusions are as follows:
\begin{enumerate}
    \item We demonstrate that it is possible to simulate SIDM models consisting of multiple species with unequal masses. However, the range of mass ratios that can be simulated reasonably well is limited. The larger the difference in mass, the more challenging the simulation, given the mass ratio of the numerical particles has to be equal to the physical mass ratio.
    \item Mass segregation allows for the formation of halos with a central density higher than that of their counterparts in the case of collisionless DM.
    \item The two-species model we consider here does not inevitably lead to gravothermal collapse, as known from commonly studied SIDM models. Instead of a catastrophe, it can result in a final state with a finite or at least slowly growing DM density in the centre of the halo.
\end{enumerate}

This paper is only a first step in exploring the phenomenology of an SIDM model consisting of multiple species with unequal masses. By providing the methodological foundations, it sets the foundation for studying further variations of unequal mass models in $N$-body simulations. Simulating various astrophysical objects employing two-species DM models will potentially allow us to further constrain the DM parameter space. These models can provide interesting signatures that could eventually shed light on puzzling observations, such as fairly compact substructures in galaxies. 

\begin{acknowledgements}
The authors thank Mathias Garny, Felix Kahlhoefer, Valentin Thoss, Hai-Bo Yu and Yi-Ming Zhong for helpful discussions and the anonymous referee for helpful comments that improved the paper.
They are also grateful for the support by Klaus Dolag, Geray Karademir, Tarun Kumar Jha, and Ashish Chittora. 
Moreover, they thank all participants of the Darkium SIDM Journal Club for discussions.
This work is funded by the \emph{Deutsche Forschungsgemeinschaft (DFG, German Research Foundation)} under Germany’s Excellence Strategy -- EXC-2094 ``Origins'' -- 390783311.
MSF acknowledge support by the COMPLEX project from the European Research Council (ERC) under the European Union’s Horizon 2020 research and innovation program grant agreement ERC-2019-AdG 882679.
Software: NumPy \citep{NumPy}, Matplotlib \citep{Matplotlib}, SciPy \citep{SciPy}.
\end{acknowledgements}

\bibliographystyle{aa}
\bibliography{bib.bib}

\begin{appendix}
\section{Stability test} \label{sec:stability_test}
\begin{figure*}[t]
    \onecolumn
    \centering
    \includegraphics[width=0.9\textwidth]{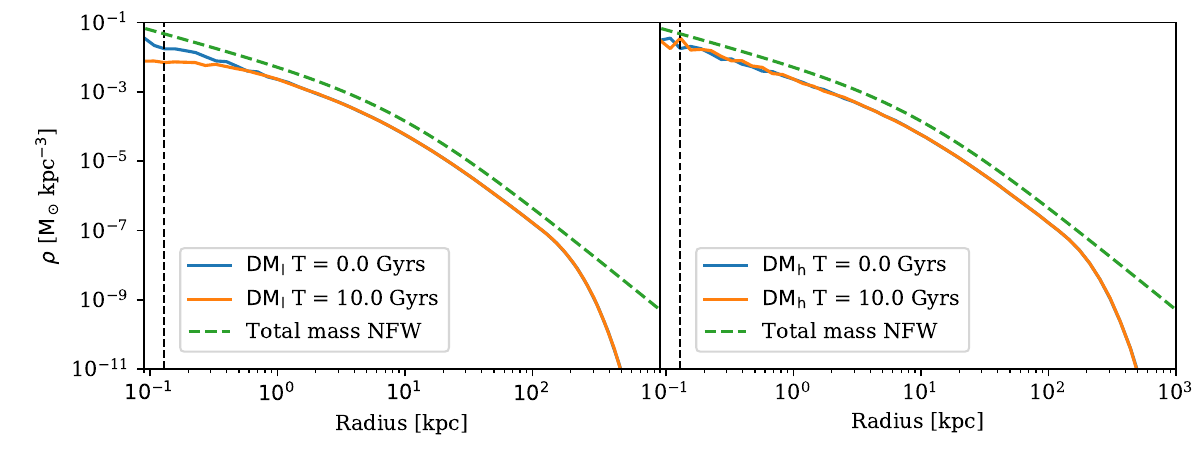}
    \caption{Density profile for collisionless DM for the $f = 0.2$, $\mu = 4$ scenario. \textit{Left}: Density profile for the lighter particles. \textit{Right}: Density profile for the heavier particles. The dashed black line indicates the softening length. The density profiles stay fairly unchanged, proving the stability of the halo. There are only minor signs of mass segregation in the inner regions of the halo.}
    \label{fig:CDM_stability}
\end{figure*}

\begin{minipage}{\columnwidth-0.25mm}
Here, we show the simulation results carried out to confirm the stability of the IC produced by \textsc{SpherIC}. We take one of the most challenging set-ups that we have simulated, namely the $f = 0.2$, $\mu = 4$ scenario. The isolated halo generated is run with CDM only, for a period of 10~Gyr. The density profiles for both species are shown individually, along with the total density from an NFW profile in Fig.~\ref{fig:CDM_stability}. We observe a very small density core formation for the lighter species, which can be attributed to gravitational relaxation leading to mass segregation in low-resolution simulations~\citep[e.g.][]{Ludlow_2019,Ludow_2023}.
We note that this gravitational relaxation is a numerical artefact. The true relaxation time due to gravitational interactions is much longer because the physical DM particles are more numerous and lighter than our simulation particles. Nevertheless, the density profiles at larger radii show no significant change for both species over time.
This is because our simulation has a sufficiently high resolution, i.e.\ enough numerical particles with a small enough mass to keep the artificial relaxation for a long enough time. As a result, the non-gravitational interactions determine the evolution of the simulations presented in Sect.~\ref{sec:isolated_halo}, not numerical artefacts.
\end{minipage}

\section{Energy conservation} \label{sec:energy_cons}
\begin{minipage}{0.5\columnwidth-2.25mm}
  In this section, we check that the simulations we have run do not violate energy conservation. Figure~\ref{fig:energy cons} shows the total energy of the simulation at different times as a fraction of the absolute energy at the initial time. We observe that for all our simulations, the energy remains very close to the initial energy for most of the simulated time period. At stages with strongly increased density, the error in simulations due to numerical issues also increases, as found in studies of the gravothermal SIDM collapse, as per the discussion in Sect.~\ref{sec:discussion}.
\end{minipage}
\hspace{2.5mm}
\begin{minipage}{0.5\columnwidth-2.25mm}
    \vspace{-15.5cm}
  \begin{figure}[H]
    \centering
    \includegraphics[width=\textwidth]{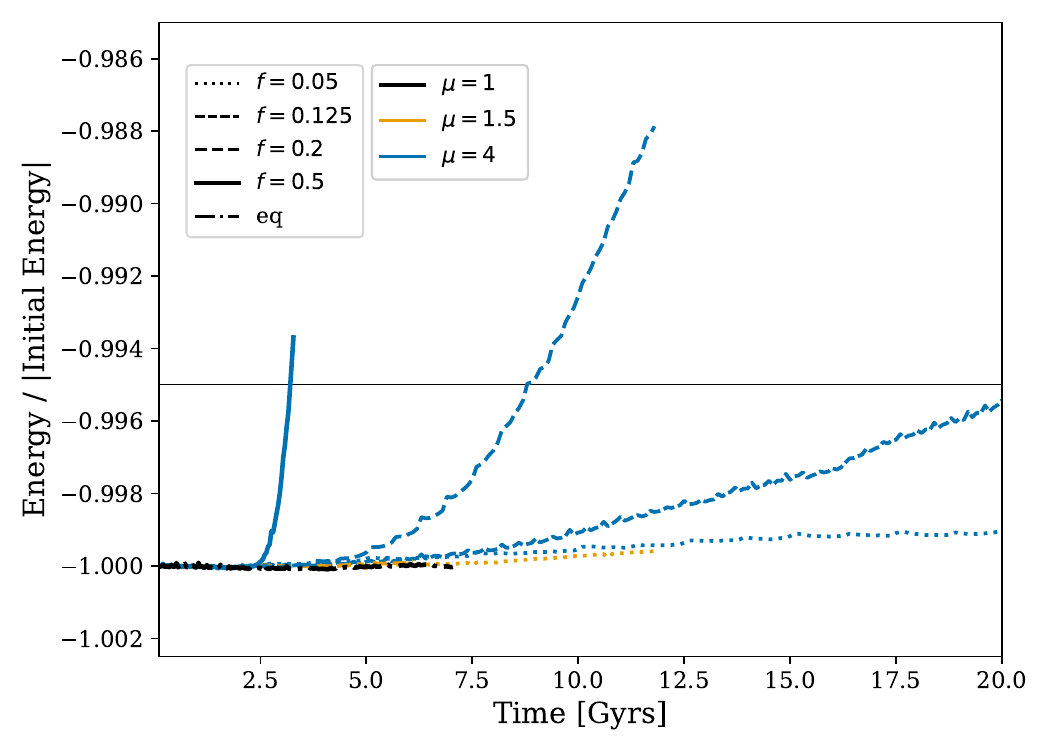}
    \caption{Energy conservation for several simulations. The total energy of the simulated particles as a fraction of the initial energy is shown as a function of time. In all other figures, we only include simulation data that does not exceed an energy error margin of 0.5\%.}
    \label{fig:energy cons}
\end{figure}
\end{minipage}

\end{appendix}

\end{document}